\documentclass[journal,letterpaper,twocolumn]{IEEEtran}
\usepackage{graphicx}
\usepackage{epstopdf}
\usepackage{CJK}
\usepackage{cite}
\usepackage{amsmath}
\usepackage[linesnumbered, ruled]{algorithm2e}
\usepackage{color}
\usepackage{amssymb}
\usepackage{multirow}
\usepackage{subfigure}
\begin{document}

\title{Path Planning for the Dynamic UAV-Aided Wireless Systems using Monte Carlo Tree Search}

\author{{Yuwen Qian, \(\textit{Member, IEEE}\), Kexin Sheng, Chuan Ma, \(\textit{Member, IEEE}\), Jun Li, \(\textit{Senior Member, IEEE}\), \\
Ming Ding, \(\textit{Senior Member, IEEE}\), and Mahbub Hassan, \(\textit{Senior Member, IEEE}\)}
\thanks{Y. Qian, K. Sheng, C. Ma, and J. Li are with School of Electronic and Optical Engineering, Nanjing University of Science and Technology, Nanjing, China. E-mail:chuan.ma@njust.edu.cn. M. Ding is with Cyber-Physical Systems, Data61, Sydney, Australia. Mahbub Hassan is with the School of Computer Science and Engineering, University of New South Wales (UNSW), Sydney, Australia.}}

\maketitle


\begin{abstract}

For UAV-aided wireless systems, online path planning attracts much attention recently. To better adapt to the real-time dynamic environment, we, for the first time, propose a Monte Carlo Tree Search (MCTS)-based path planning scheme. In details, we consider a single UAV acts as a mobile server to provide computation tasks offloading services for a set of mobile users on the ground, where the movement of ground users follows a Random Way Point model. Our model aims at maximizing the average throughput under energy consumption and user fairness constraints, and the proposed timesaving MCTS algorithm can further improve the performance. Simulation results show that the proposed algorithm achieves a larger average throughput and a faster convergence performance compared with the baseline algorithms of Q-learning and Deep Q-Network.
\end{abstract}




\IEEEpeerreviewmaketitle

\section{Introduction}

Due to the high mobility and low cost, unmanned aerial vehicles (UAVs) have been found a wide range of applications in the communication field during the past few decades. For example, UAV-aided wireless communications can provide seamless connection to devices without communication infrastructure coverage due to severe building shadowing~\cite{8675384}. {\color{black}In~\cite{9442809}, the authors used UAVs to cover users in wireless networks, and proposed an efficient clustering method to determine the optimal UAV positions.} In UAV-mounted mobile edge computing (MEC), the UAV can be employed as a mobile edge server to compute the tasks offloaded by users.

One important application in the UAV systems is path planning~\cite{Zhang2018UAV}. Authors in~\cite{8913466} proposed a UAV-aided data collection scheme with an objection to optimize the UAV's trajectory, velocity and data links in a 3-D space. In~\cite{9173532}, a nonstationary geometry-based stochastic model for UAV-to-ground channels was studied, which supported the channel generation through arbitrary 3-D trajectories. Authors in~\cite{8895810} studied a UAV-enabled edge-cloud system with an aim to jointly optimize the allocation of resources and the UAV trajectory in a 3-D space. In addition, a UAV path planning problem is usually limited by multiple constraints like energy consumption limits and quality of service (QoS). The authors in~\cite{2018Throughput} jointly optimized the communication scheduling, power allocation and UAV trajectory. In~\cite{8870206}, the authors jointly optimized the UAV trajectory and wireless resource allocation under the users' maximum power constraints. In~\cite{2018Mobile}, the authors aimed at minimizing the energy consumption while satisfying the quality of service requirements, and used a convex optimization method to solve this problem. However, due to the limitation of the traditional path planning methods, i.e., the A-star algorithm in~\cite{2014AStar}, solving the problems with multiple constraints becomes challenging. Therefore, researchers have applied the reinforcement learning (RL) technique to UAV path planning~\cite{9024685}.


In the RL, an agent interacts with the environment by evaluating the value of selected actions and updating the reward generated by environment. This information is then back propagated to select the next state and action for UAVs. Authors in~\cite{2017Q} designed a UAV path learning and obstacle avoidance method based on Q-learning (QL) algorithm, which uses QL to allow UAV learning environment continuously. QL-based schemes need to store all state-action pairs in a Q-table, and they will meet limitations to deal with the high-dimensional states. In the light of that, the authors in~\cite{2019Towards} employed the Dueling Double Deep Q-Networks algorithm that takes a set of situation maps as input to approximate the Q-function, which can alleviate the curse of high dimensionality with a certain performance loss.

Authors in~\cite{2018Energy} designed UAV trajectories in a static environment, but the design tends to be nontrivial when the environment becomes dynamic, such as the time-varying user locations, variable user task requirements, and unstable UAV-user channel states. In details, the task requirements of the user in each time interval will influence the migration throughput, and the variety of mobile users will alter the communication channel state as well as the energy consumption. More importantly, the instability of the channel state will affect the task transmission quality and the flying trajectory. In this case, path planning in dynamic environment with multiple varying conditions attracts research attentions. {\color{black}According to~\cite{lowe2020multiagent}, conventional RL methods like QL~\cite{2017Q} and Deep Q-Networks~\cite{2019Towards} (DQN) perform worse in uncertain environments. Besides, they have a comparable high time cost on dealing with a massive space of state-action pairs,} which will decrease the efficiency of path planning and cannot be adapted to the dynamic environment directly. Different from the traditional RL methods, Monte Carlo Tree Search (MCTS)~\cite{2012A} can make trade-offs in unknown scenarios and generate heuristic solutions, which leads to a potential for dealing with the relationship of states and actions with a low time complexity.

How to design the UAV trajectory efficiently and instantly in dynamic environment still remains to be a challenging task. Motivated by this research gap, we, for the first time, introduce MCTS to UAV path planning. We consider a typical UAV-aided wireless communication scenario, where the location of users, communication channel state, and the user task requirements are all time-varying. The proposed algorithm can well capture the dynamics of the system by the real-time playout, and achieve an incremental performance compared to other RL based algorithms with a comparable low complexity.

The rest of the paper is organized as follows. In Section~\ref{sec2}, we present the system model and energy consumption model, as well as the problem formulation. Section~\ref{sec3} shows our proposed algorithm including the algorithm procedure and time complexity analysis. Besides, we provide an improved algorithm with a lower time complexity. In Section~\ref{sec4}, we provide the experimental results, followed by conclusions in Section~\ref{sec5}.
\section{System Model And Problem Formulation}
\label{sec2}
\subsection{System Model}

\begin{figure}
\centering
\includegraphics[height=4.64cm,width=7.18cm]{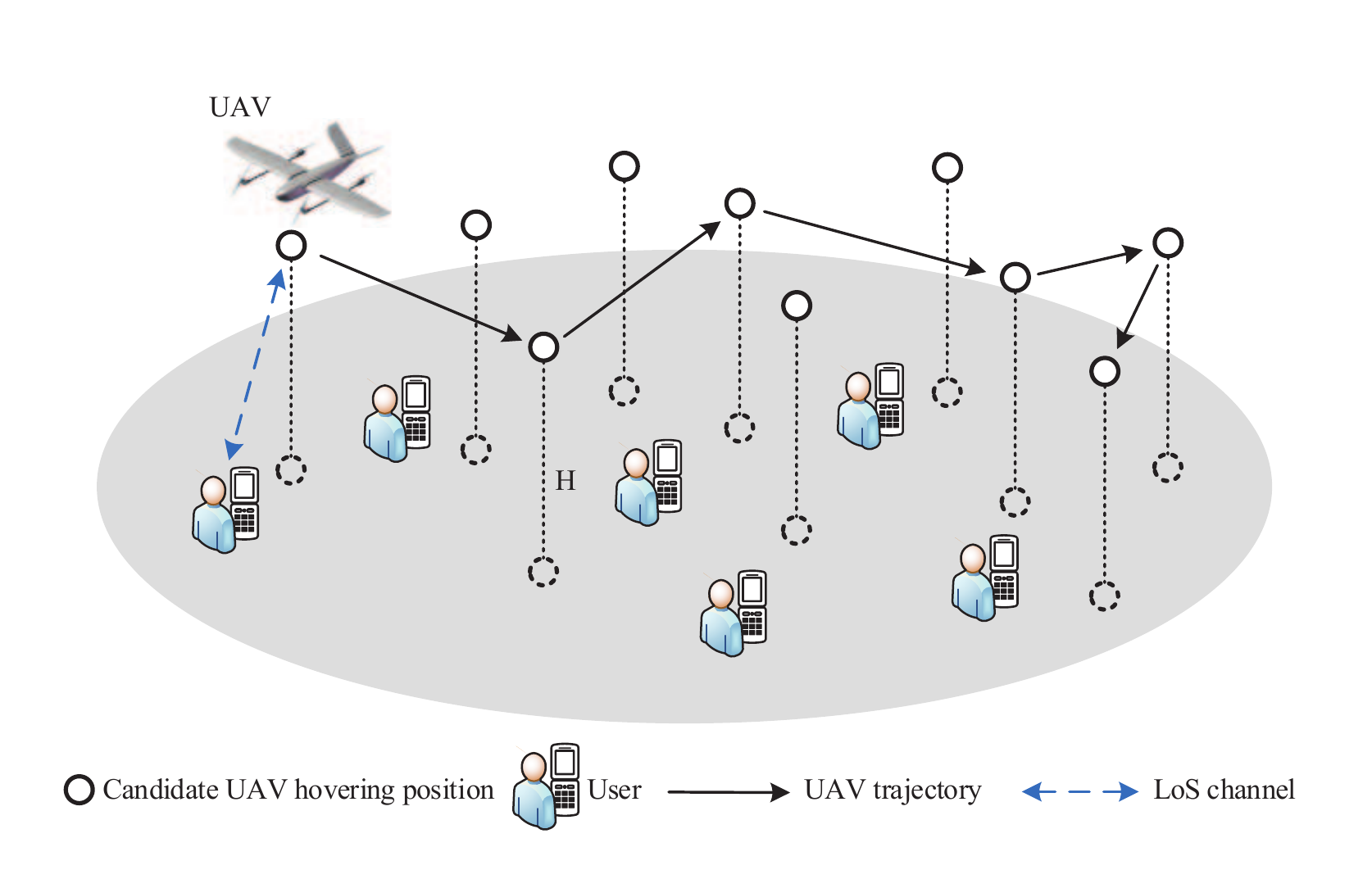}
\caption{A typical wireless communication scenario where a single UAV provides services for multiple ground users.}
\label{fig_scenario}
\end{figure}

As shown in Fig~\ref{fig_scenario}, we consider a typical wireless communication scenario where a single UAV provides services for multiple users on the ground. The users are distributed randomly with an original location \(U_{i}(0)=[x_{i}(0), y_{i}(0), 0]\) in this region, where $i$ is the index of users and the third dimension represents the height of users. Meanwhile, there are $K$ fixed hovering points above these users~\cite{9234110}, with the three-dimensional coordinate \(D_{k}=[x_{k}, y_{k}, H]\).\footnote{{\color{black}We have also modified the UAV trajectory into a 3-D one. The newly added results can be found in Appendix B.}} Following the same assumption with $D_{\mathrm{uav}}(t)=[x_{\mathrm{uav}}(t), y_{\mathrm{uav}}(t), H]$, the UAV can only hover at these fixed points and provide task computation service for the users in this area. In addition, we consider the UAV flies horizontally at a fixed altitude \(H\).


We suppose the total service time $T$ is divided into $N$ unequal time intervals due to different flying and hovering times, where \(t=0, 1, 2,..., N-1\). Within each time interval, the UAV flies from one UAV hovering position to any other candidate fixed point and hovers to serve for a certain user. The UAV tends to serve the nearest user. Moreover, we assume that at the beginning of each time slot, all users will send the task request with their location information to the UAV. Thus, the UAV can obtain the location information of users. Users are randomly distributed when $t=0$, and the location of users will change at the beginning of the next time interval following the Random Way Point (RWP) model~\cite{2011Capacity}. That is, users move to a destination with a random speed and destinations are randomly distributed over the given area. The location updates can be depicted by the following expression:
\begin{equation}
x_{i}(t)=
\begin{cases}
x_{i}(t-1)+\alpha_1,& \mathrm{if}\ 0\leq x_{i}(t-1)+\alpha_1 \leq R;\\
R,& \mathrm{if}\ x_{i}(t-1)+\alpha_1>R;\\
0,& \mathrm{if}\ x_{i}(t-1)+\alpha_1<0.
\end{cases}
\end{equation}
\begin{equation}
y_{i}(t)=
\begin{cases}
y_{i}(t-1)+\alpha_2,& \mathrm{if}\ 0\leq y_{i}(t-1)+\alpha_2 \leq R;\\
R,& \mathrm{if}\ y_{i}(t-1)+\alpha_2>R;\\
0,& \mathrm{if}\ y_{i}(t-1)+\alpha_2<0,
\end{cases}
\end{equation}
where \(\alpha_1\), \(\alpha_2\) are random numbers between -$\varepsilon$ and $\varepsilon$, and $\varepsilon$ represents the moving range. $R$ presents the upper bound of the user location. To ensure the service quality of each user, UAV can only serve one user in a time interval\footnote{{\color{black}If we consider that the UAV serves multiple users in a time interval, it may be necessary to optimize the bandwidth allocation and resource allocation. Thus, how to design the UAV trajectory in such a scenario is a completely different problem, and we will consider it as one of our future research work.}}, and manages to meet the service requirement of all users during the flight. The user fairness constraint during one period is defined as follows:
\begin{equation}
b_{i}(t)=
\begin{cases}
1,& \mathrm{if}\sum\limits_{n=0}^{t-1} b_i(n)=0\ \mathrm{or}\sum\limits_{n=0}^{t-1} b_i(n)=1\ \&\ u_i(t)>\beta;\\
0,& \mathrm{otherwise},
\end{cases}
\label{eq_user}
\end{equation}
where \(u_i(t)\) is the number of offloaded tasks, $b_i$ is a binary variable that indicates whether the user is being served or not, and $\beta$ is a given task threshold. If the user has been served before and the second task requirement is larger than $\beta$, it will be served.

The distance between UAV and users can be expressed as
\begin{equation}
d(t)=\sqrt{H^{2}+\lVert D_{\mathrm{uav}}(t)-U_{i}(t)\lVert ^{2}}.
\end{equation}
For simplicity, we suppose the communication link between the UAV and users is line-of-sight (LoS) link, and a a combination of LoS and NLoS channel model can be considered as a further work. Furthermore, the channel transmission quality is assumed to relate to the distance between UAV and users~\cite{2017Joint}. In this way, the channel power gain between the user and the UAV follows the free-space path loss model, which can be expressed as:
\begin{equation}
g_{i,k}(t)=\frac{\rho_0}{H^{2}+\lVert D_{\mathrm{uav}}(t)-U_{i}(t)\lVert ^{2}},
\end{equation}
where \(\rho_0\) denotes the channel gain at the reference distance 1 meter.

\subsection{Energy Consumption}

UAV follows the energy consumption constraint during the whole flight process. Let $\theta$ be a given battery threshold, and the UAV will not stop serving the area until its battery percentage level falls below $\theta$. There are three main types of energy conservation:

\subsubsection{Hovering Energy Consumption}

When UAV is located at the $k$-th hovering point, it will wait for the $i$-th user uploading tasks and consume the corresponding energy when hovering. According to~\cite{2017Joint}, the uploading rate (bits/s) from the $i$-th user to the UAV within the $t$-th time interval is given by
$R_{i,k}(t)=\log_2(1+\frac{P_\mathrm{u}g_{i,k}(t)}{\sigma^{2}})$,
where \(P_\mathrm{u}\) denotes the uploading transmission power of the associated user, and \(\sigma^{2}\) is the power of additive white Gaussian noise. The bits of offloaded tasks can be expressed as
$S(t)=u_i(t) B_\mathrm{t}$,
where \(B_\mathrm{t}\) is the amount of bits for one task. Let \(P_\mathrm{h}\) be the UAV hovering power, the hovering energy consumption \(e_{\mathrm{h}}(t)\) is given as~\cite{9044434} $e_{\mathrm{h}}(t)=P_\mathrm{h} \frac{S(t)}{R_{i,k}(t)}$.

\subsubsection{Computation Energy Consumption}

The computation energy for calculating offloaded tasks from each user can be described as
\begin{equation}
e_{\textrm{c}}(t)=\gamma_\mathrm{c}CS(t)f_\mathrm{c}^2,
\end{equation}
where \(\gamma_\mathrm{c}\) is the effective switched capacitance, \(C\) is the number of CPU cycles per input bit needed for computing, and \(f_\mathrm{c}\) is the CPU frequency.

\subsubsection{Flying Energy Consumption}
Assuming that UAV flying speed \(v(t)\) is a constant, the energy consumed due to flying to another hovering point can be defined as
\begin{equation}
e_{\mathrm{f1}}(t)=\kappa v^2(t),
\end{equation}
where \(\kappa=0.5M\delta_t\), \(\delta_t=\frac{\lVert D_{\mathrm{uav}}(t)-D_{\mathrm{uav}}(t-1) \lVert  }{v(t)} \) and \(M\) is mass of the UAV~\cite{2018Mobile}. Moreover, the UAV also has a propulsion energy consumption because of the acceleration from a stationary state to a flying state, which can be modeled as:
\begin{equation}
e_{\mathrm{f2}}(t)=\kappa_1 v^3(t)+\frac{\kappa_2}{v(t)}(1+\frac{a_{\mathrm{uav}}^2}{g^2}),
\end{equation}
where $\kappa_1$, $\kappa_2$ are all system parameters, $a_{\mathrm{uav}}$ denotes the acceleration of UAV and $g$ is the gravitational acceleration. Thus the flying energy consumption is $e_{\mathrm{f}}(t)=e_{\mathrm{f1}}(t)+e_{\mathrm{f2}}(t)$.

Thus, the total energy consumption within one single time interval is \(w(t)=e_\mathrm{h}(t)+e_\mathrm{c}(t)+e_\mathrm{f}(t)\), while the energy consumption during the whole flight is denoted by \(W(t)=\sum\limits_{t=0}^T [e_\mathrm{h}(t)+e_\mathrm{c}(t)+e_\mathrm{f}(t)]\).

\subsection{Problem Formulation}
\label{sec_pro}
We define the average throughput as the average offloaded bits from users during $T$ time intervals, and our goal is to maximize the average throughput. Toward this end, we design the UAV trajectory subject to the energy conservation constraint and user fairness constraint. In this way, the problem can be formulated as:

\begin{subequations}
\begin{alignat}{2}
\max\quad & \frac{\sum\limits_{t=0}^{T-1}S(t)}{T}, &{}& \tag{9a} \label{eq_11}\\
\mbox{s.t.}\quad
&\sum\limits_{t=0}^{T-1} e_{\mathrm{f}}(t)+e_{\mathrm{h}}(t)+e_{\mathrm{c}}(t) \leq (1-\theta)E_0, &{}& \tag{9b} \label{eq_22}\\
&\sum\limits_{t=0}^{T-1}b_i(t)\leq 2,\ \mathrm{and}\ (\ref{eq_user}), &{}& \tag{9c} \label{eq_33}
\end{alignat}
\end{subequations}
where (\ref{eq_11}) represents the average offloaded bits from users, (\ref{eq_22}) indicates that total energy consumption is no more than $(1-\theta)$ of the original UAV battery $E_0$ and (\ref{eq_33}) guarantees the service fairness of users. To solve the problem (\ref{eq_11}) directly, the complexity tends to be $\mathcal{O}(K^I)$, where $K$ denotes the number of hovering points, and $I$ is the number of users, and it is an NP-hard problem which can not be solved in the polynomial time. Thus, we will employ the MCTS-based algorithm in this next section, which can well capture the dynamics of the environment with a comparable low complexity.

\section{Proposed Algorithm}
\label{sec3}

In this section, we introduce the basic algorithm of MCTS and the MCTS-based UAV path planning scheme. In our proposed algorithm, UAV selects the next action according to current state and then receives feedback from environment. The UAV will adjust actions according to the reward. Particularly, MCTS attaches the data information to the corresponding tree node for easy access and data operations, which is fit for real-time path planning and can accelerate training process.
\subsection{Monte Carlo Tree Search}

Monte Carlo Tree Search (MCTS) is a method to find optimal decisions in a given domain by randomly sampling in the decision space~\cite{2012A}. Different from traditional reinforcement learning, it can make decisions in real-time according to the feedback of environment instead of relying on the results of off-line training, which is more suitable for practical situation. MCTS constructs a game tree by evaluating actions in different observations and finally finds the best strategy.
The tree consists of numerous tree nodes and conjunctional lines, and each node represents a state in current situation.
In the proposed algorithm, the top-down search tree represents a decentralization of UAV flying states.


According to~\cite{2012A}, the $n$-th node contains the following information: {\color{black}UAV state $s(n)$, action $a(n)$, quality value $Q(n)$, visiting count $N(n)$, parent node $n_p$ and child node $n_c$. Also, state $s(n)$ is comprised of UAV position, user locations and UAV battery $E$.} Each level of the search tree represents the current time interval of the UAV. Four steps are applied during each iteration:
\subsubsection{Selection}
Starting from the root node, Upper Confidence Bound for Tree (UCT) algorithm is adopted to select the most appropriate child node until reaching a node which is not fully expanded or a leaf node. That is, UAV will continue flying to another hovering point and provides service for a certain user. UCT can well balance the relationship between exploration and exploitation, which is expressed as:
\begin{equation} \label{update}
\mathrm{UCT}=\mathop{\mathrm{arg max}}\limits_{n_0'\in \mathrm{child\ of}\ n_0}\frac{Q(n_0')}{N(n_0')}+c\sqrt{\frac{2\ln N(n_0)}{N(n_0')}},
\end{equation}
where the constant \(C\) is a trade-off factor between exploration and exploitation, \(Q(n_0')\) is sum of reward value, \(N(n_0')\) denotes the visit count of the child node and \(N(n_0)\) denotes the visit count of the parent node.
\subsubsection{Expansion}
Once the current node has nonterminal state and unexpanded children, a new child node will be created and initialized from the current node~\cite{2012A}. Since the other child nodes of the current node contain existing actions, the next access point the UAV selects should be the one has not been selected in the current state.

\subsubsection{Playout}
Execute random policy starting from the newly expanded node until a terminal state is reached. 
That is, UAV will continue flying from current access points and the actions of UAV are chosen randomly during the whole process. Once a critically low energy state is reached, the UAV will drop out from providing service and return to its base for battery charging. Note that there is no new tree node being created, instead, the corresponding virtual states are generated. The objective of this step is to update the statistics of the tree node, which contributes to trajectory optimization and better action selection of the UAV.
\subsubsection{Backpropagation}
Calculate the reward value generated during simulation. The reward function in a single time interval is defined as
\begin{equation}
q_i=\frac{u_i(t)}{u_{\mathrm{max}}}-\frac{1}{w_{\mathrm{max}}}[e_\mathrm{h}(t)+e_\mathrm{c}(t)+e_\mathrm{f}(t)],
\end{equation}
where $u_{\mathrm{max}}$ and $w_{\mathrm{max}}$ represent the maximum of task numbers and energy consumption, respectively. As the mobility of users will affect the selection of the served user, and the task requirement of the user can affect the system average throughput and average reward, so to maximize the average throughput, the reward function should grow as the throughput increases. On the other hand, the UAV has an energy consumption constraint, so the energy consumption should have a negative impact on the reward function.

In addition, the average reward value in the random policy is given by
$R=\sum\limits_{i=t+1}^T \frac{q_i}{T-t}$.
The state of each node along with the path will be updated from the current node to the root node, including their visit count and quality value.

\begin{algorithm}
\label{alg_alg1}
\caption{MCTS-based path planning Scheme of Maximizing Average Throughput}

\textbf{initialize:} $n_p=$ NONE and $n_c=$ NONE; \label{line1}\\
\textbf{initialize:} $N(n)=$ 0 and $Q(n)=$ 0; \label{line2}\\
\textbf{initialize:} $episode = 0$ and $E=E_0$; \label{line3}\\
\Repeat{$n\ \mathrm{is\ a\ terminal\ state}$}
{
\For{episode $=$ 0 to $N_e$ (total training times)}
{
\While{battery $E > 0$}
{ \label{line6}
\eIf{node $n$ is fully expanded}
{ \label{line7}
UAV executes action $a(n)$;\\
choose a user to serve;\\
$n \gets \mathop{\mathrm{\mathrm{arg max}}}\limits_{n'\in \mathrm{\mathrm{child\ of\ }}n} \frac{Q(n')}{N(n')}+c\sqrt{\frac{2\ln N(n)}{N(n')}}$; \label{line9}\\
}
{
add a new child node $n_1$ to $n$; \label{line11}\\
choose an action $a_1$ and $a(n_1)=a_1$; \label{line13}\\
\textbf{break}
}
$E \gets E-\frac{1}{w_{\mathrm{max}}}[e_\mathrm{h}(t)+e_\mathrm{c}(t)+e_\mathrm{f}(t)]$;\\
}
\label{line17}
\While{battery $E > 0$}
{ \label{line18}
choose an action $a'$ randomly;\\
UAV executes action $a'$;\\
choose a user to serve;\\
$E \gets E-\frac{1}{w_{\mathrm{max}}}[e_\mathrm{h}(t)+e_\mathrm{c}(t)+e_\mathrm{f}(t)]$;\\
} \label{line23}
calculate reward $R$ \label{line24}

\While{$n_1$ is not null}
{ \label{line25}
$N(n_1) \gets N(n_1) + 1, Q(n_1) \gets Q(n_1)+R$;\\
$n_1 \gets$ parent of $n_1$;\\
} \label{line28}
}
$n \gets \mathop{\mathrm{\mathrm{arg max}}}\limits_{n'\in \mathrm{\mathrm{child\ of\ }}n} \frac{Q(n')}{N(n')}+c\sqrt{\frac{2\ln N(n)}{N(n')}}$; \label{line31}\\
}
\end{algorithm}

The detailed description of MCTS-based path planning scheme is given in Algorithm~\ref{alg_alg1}. In lines~\ref{line1}-\ref{line3}, the statistics of the root node is initialized. In the beginning, the UAV is located at initial position while the search tree starts from the root node. If current node is fully expanded, the next action is guided by the UCT algorithm that the best child node is chosen with the maximum UCT value and UAV will execute a corresponding action to serve a certain user, which is described in lines~\ref{line7}-\ref{line9}. Assuming that one user has been served once before, for equality, it can be served twice only when task requirement is more than $\beta$. From lines~\ref{line11}-\ref{line13}, a new node is expanded and UAV flies to an unvisited access point. In lines~\ref{line18}-\ref{line23}, the UAV will follow a random flying trajectory on the basis of random policy. In lines~\ref{line24}-\ref{line28}, the generated reward will be calculated and be back up along the path. Then the best child node is selected to start the next iteration in line~\ref{line31}.



\subsection{Time Complexity Analysis}
\label{time}

In this subsection, we analysis the time complexity of the proposed algorithm. As can be observed, Algorithm~\ref{alg_alg1} has five loop bodies. Let the level of the tree be $m$ and the times of training episodes be $n$. Assume that the number of iterations for selecting child nodes and expanding a new node is $x$ in lines~\ref{line6}-\ref{line17}, the loops in lines~\ref{line18}-\ref{line23} and lines~\ref{line25}-\ref{line28} have $(m-x)$ and $x$ iterations, respectively. Thus the total number of iterations in one episode is $x+(m-x)+x=m+x$. The overall time complexity for completing once tree search is $\mathcal{O}(m+x)$. Consider the whole search process, the time complexity is defined as $\mathcal{O}(n(m+x))$ in one single flight round and $\mathcal{O}(mn(m+x))$ in $m$ flight rounds. When the tree is not completely built, the layer of the tree is less than $m$, hence we have $x<m$ and the complexity is lower than $\mathcal{O}(2m^2n)$. So in the worst case when $x=m$, the time complexity is $\mathcal{O}(2m^2n)$.



\subsection{Timesaving-MCTS algorithm}

As the number of training episodes increases, the search tree improves gradually. Redundant training may not enhance the performance of algorithm significantly and will waste time. Therefore, we have improved our proposed algorithm as follows. Assume that the level of search tree is $m$ and training episodes are $n$, and the number of iterations of selection and expansion is $x$. Iteration times will decrease $\frac{n}{m}$ as the tree layer increases. For instance, iteration will be executed $n(m+x)$ times when the search starts from the first layer and $(n-\frac{n}{m})(m+x)$ times from the second layer. In this way, the total iteration number during the flight is calculated as

\begin{small}
\begin{equation}
\begin{split}
(m+x)(n+...+n-\frac{(m-1)n}{m})
=(m+x)(\frac{mn}{2}+\frac{n}{2}),
\end{split}
\end{equation}
\end{small}
and the time complexity can be expressed as $\mathcal{O}((m+x)(\frac{mn}{2}+\frac{n}{2}))$. Considering the worst case when $x=m$, the time complexity becomes $\mathcal{O}(m^2n+mn)=\mathcal{O}(m^2n)$. Compared to the time complexity of MCTS in Section~\ref{time}, the complexity of TS-MCTS is reduced by 50\%, which demonstrates the effectiveness of the improved algorithm.



\section{Experimental Results}
\label{sec4}

\begin{figure}[htbp]
\centering
\includegraphics[width=0.45\textwidth]{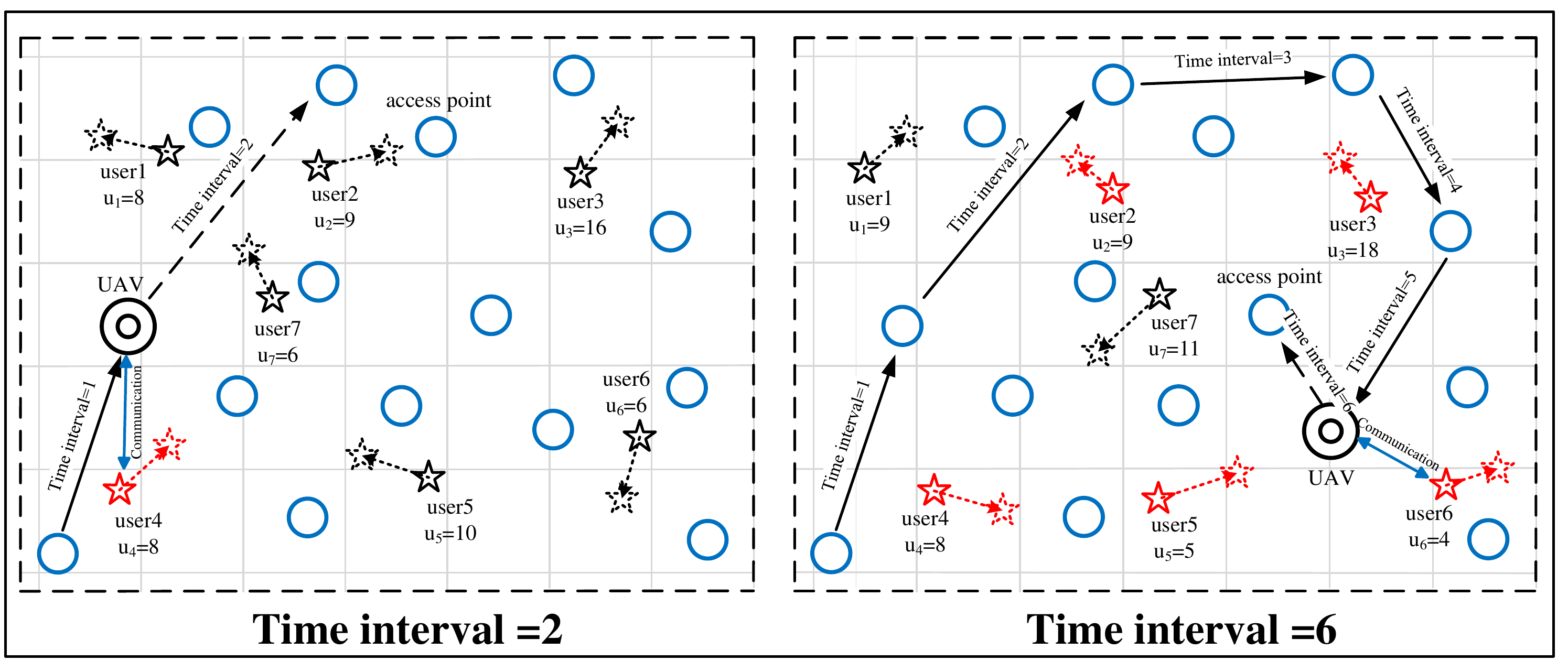}
\caption{UAV trajectory design with UAV velocity $=$ 20m/s. When $t=1$, the 4-th user and the 7-th user have a similar distance with UAV, and the UAV selects the 4-th user with a larger task requirement.}
\label{fig_trajectory}
\end{figure}

In this section, we provide experimental results to evaluate the performance of our algorithm. The number of hovering points and users are set to $K=20$ and $I=10$, respectively. We divide the users into two neat rows and randomly generate their positions, and set the distance between users to be the same. The offloaded tasks are assumed to follow a Gaussian distribution with the expectation value setting to 10 and the standard deviation setting to $\varphi=5$. The level of search tree is set to $m=10$. Other parameters values are listed as follows: $H=100$m, $M=10$kg, $\sigma^{2}=-120$dBm, $\rho_0=-50$dB, $P_\mathrm{h}=1$W, $P_\mathrm{u}=0.1$W, $B_\mathrm{t}=100$Kb, $\kappa_1=0.03$, $\kappa_2=5$, $a_{uav}=15$m/s$^2$, $g=9.8$m/s$^2$, $C=1000$, $f_c=2$GHz and $\gamma_c=$ \(10^{-28}\) F~\cite{7956189}. In addition, the task threshold $\beta$ is set to $15$ and the battery threshold $\theta$ is set to $10\%$.
In Fig.~\ref{fig_trajectory}, we show the optimal UAV trajectory using the proposed path planning scheme,
where the red stars denotes the users that have been served.

\begin{figure}[htbp]
\centering
\subfigure[UAV velocity $=$ 20m/s]{\label{fig_11}
\begin{minipage}[t]{0.48\linewidth}
\centering
\includegraphics[width=4.45cm]{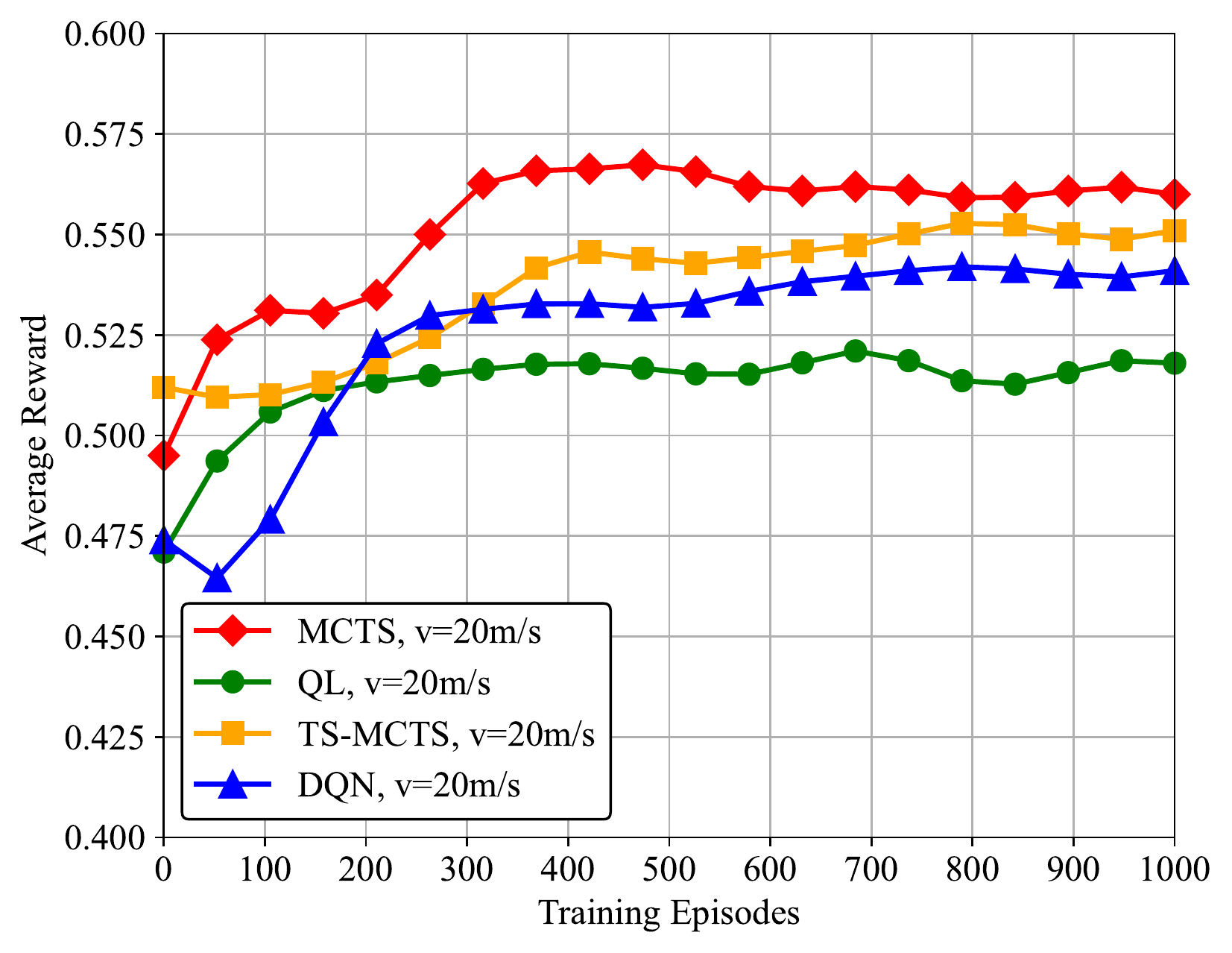}
\end{minipage}%
}%
\subfigure[UAV velocity $=$ 10m/s]{\label{fig_22}
\begin{minipage}[t]{0.48\linewidth}
\centering
\includegraphics[width=4.45cm]{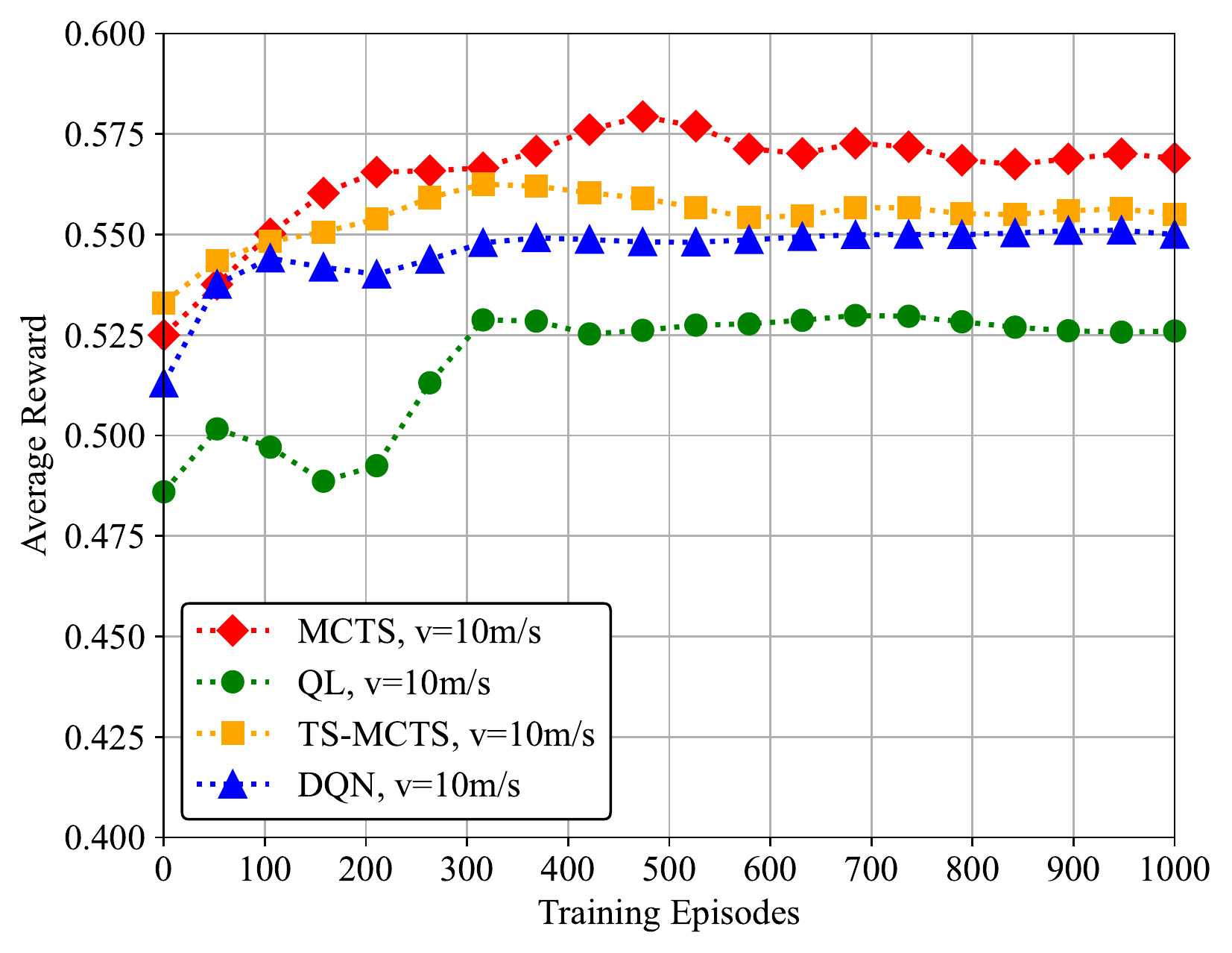}
\end{minipage}
}%
\centering
\caption{Average system reward versus training episodes with UAV velocity $=$ 20m/s and 10m/s respectively.}
\label{fig_reward}
\end{figure}

To compare with other algorithms, we employ QL~\cite{2017Q} and DQN~\cite{2019Towards} as the benchmarks, and the performance of the timesaving (TS) MCTS algorithm is also conducted. For QL, the actions are stored as a single row in the Q-table, and each action maps several different states. Every state-action pair has its own Q value. Fig.~\ref{fig_reward} illustrates the comparison of the average reward value. First, we observe that the MCTS-based algorithm can achieve the largest average reward among all algorithms. {\color{black}This is because the dynamic environment causes great increase in the number of state-action pairs, which leads to inadequate training for existing state-action pairs. Different from DQN and QL, MCTS has a stochastic simulation process, hence MCTS has the capability to capture the random changes in the environment, and performs better than QL and DQN.} Second, a lower UAV speed leads to a higher average reward for all algorithms. This is because a higher UAV speed will consume more flying energy, and hence the average reward will decrease.

\begin{figure}
\centering
\includegraphics[height=4.3cm]{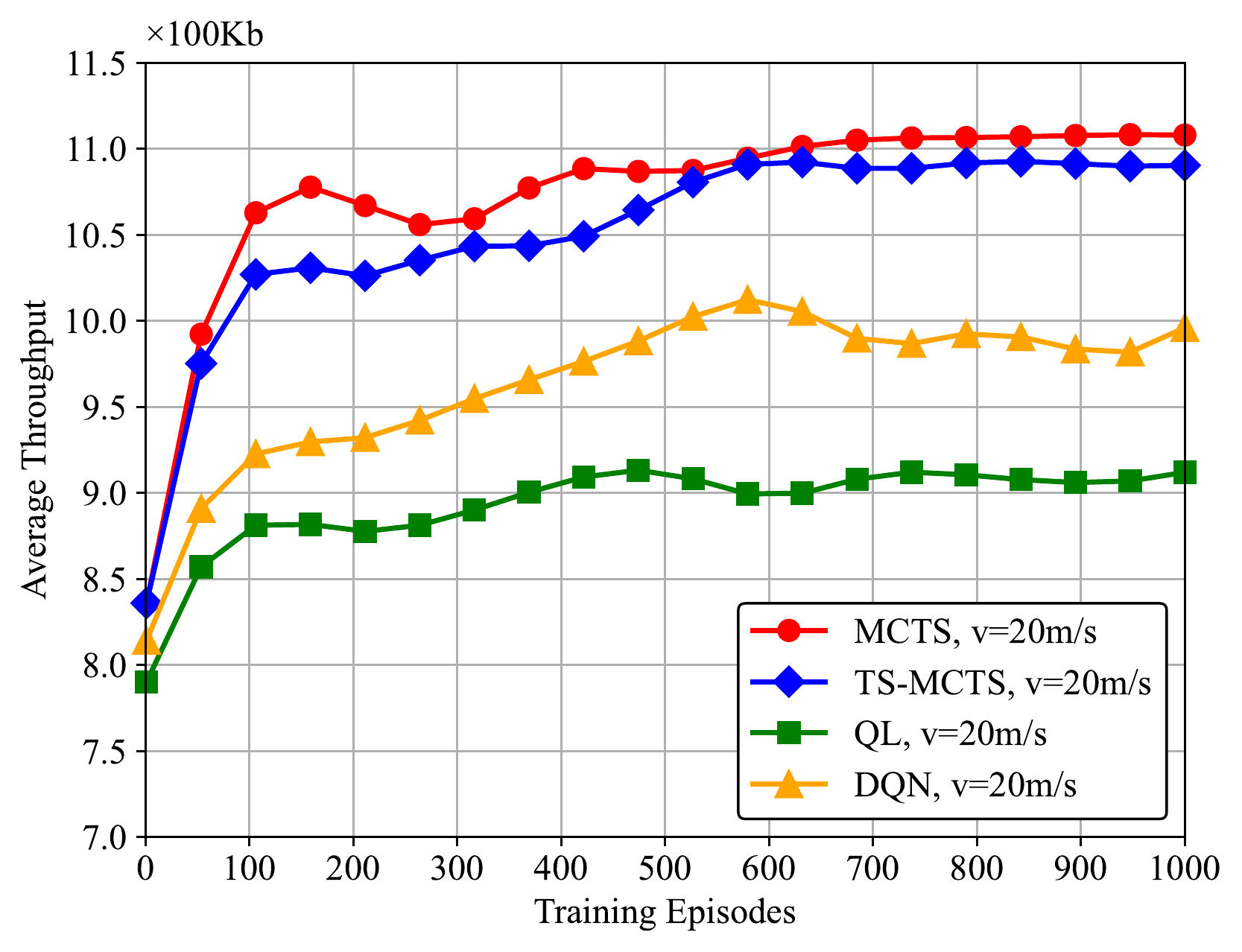}
\caption{Average throughput versus training episodes with UAV velocity $=$ 20m/s, comparing with different algorithms.}
\label{fig_throughput}
\end{figure}

Fig.~\ref{fig_throughput} depicts the average throughput with different training episodes.
For the TS-MCTS algorithm, iteration will be executed 500 times if the search starts from the first layer of the tree and be executed 450 times from the second layer when training episodes $=$ 500. As can be found in this figure, the proposed algorithm achieves the largest average throughput among these algorithms, which indicates that MCTS is more appropriate for real-time dynamic path planning than other RL methods. Then, with the increase of training episodes, the growth of the average throughput using MCTS is the largest among all the algorithms.
For example, the value of average throughput is about 8.35 when training episodes $=$ 10, and arrives at 11.1 when training episodes $=$ 1000.
In addition, although the throughput of TS-MCTS is lower than MCTS, the performance gap decreases with the increasing training epochs.

\begin{figure}
\centering
\includegraphics[height=4.3cm]{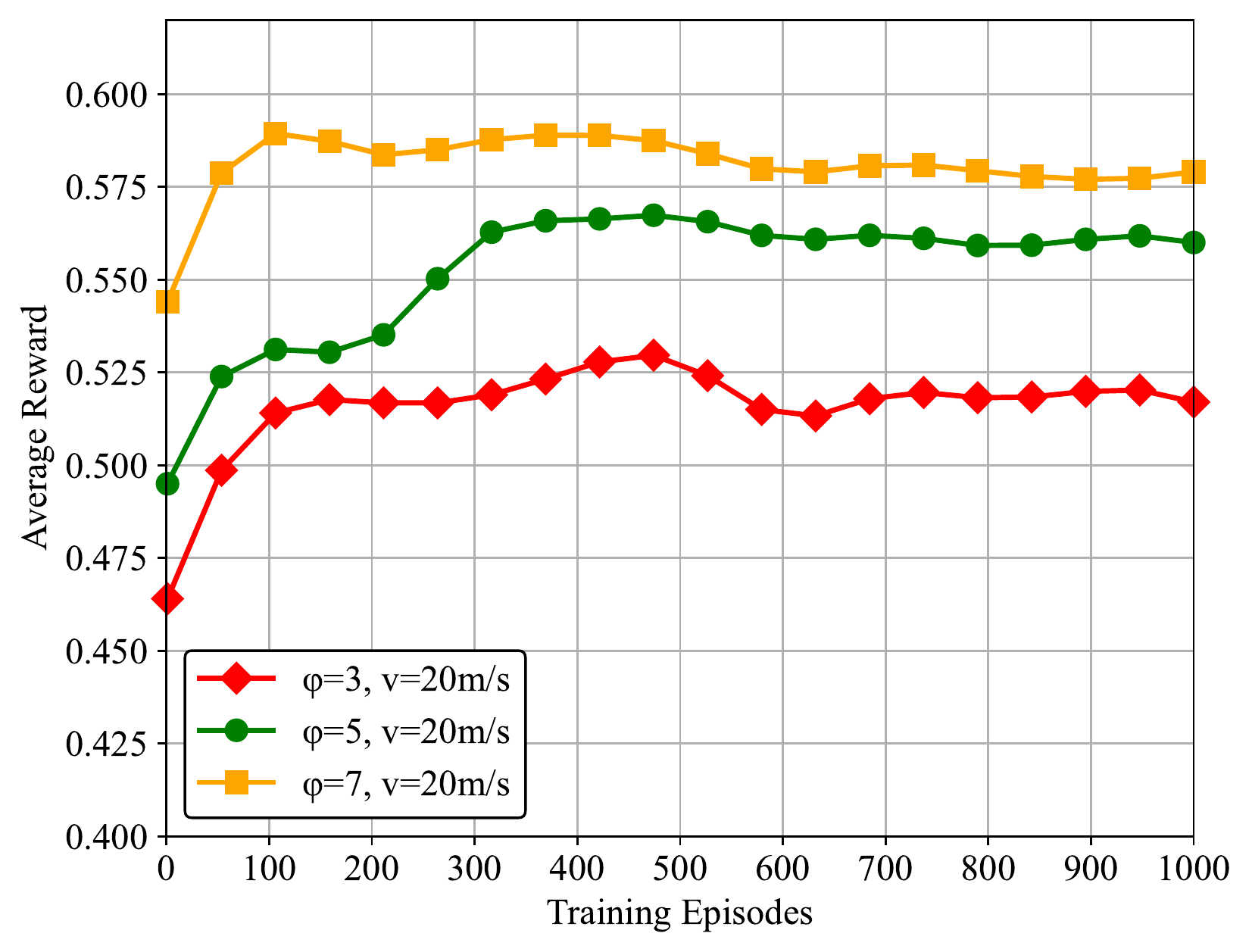}
\caption{The robustness of proposed algorithm under different task distributions.}
\label{fig_robust}
\end{figure}

In Fig.~\ref{fig_robust}, we show the robustness of our proposed algorithm under different standard deviation of the task distribution. The velocity of UAV is set to 20m/s. As can be seen from this figure, all of the curves eventually converge, and the average reward increases with a larger value of the standard deviation for task distribution.

\begin{figure}
\centering
\includegraphics[height=4.3cm]{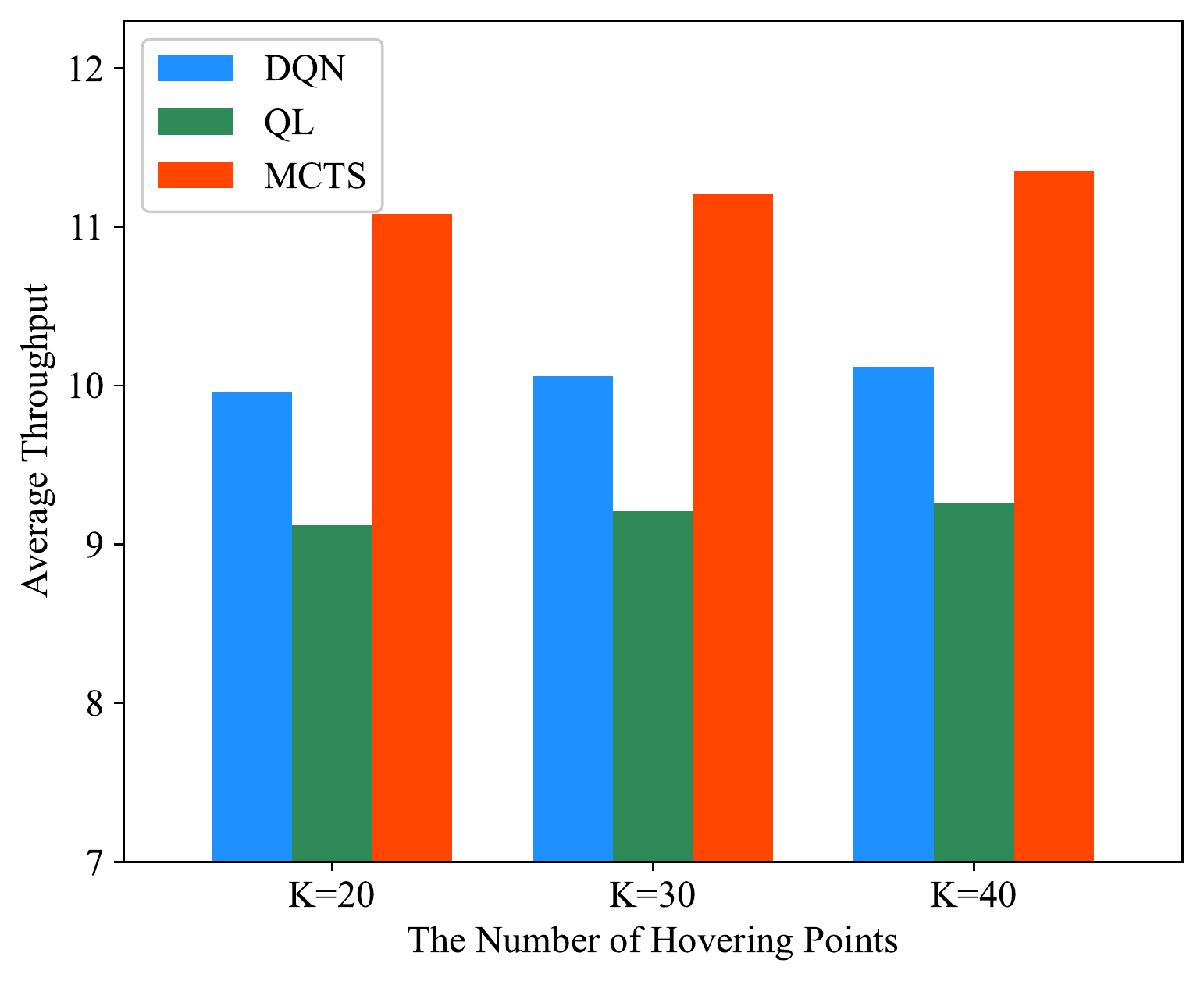}
\caption{The average throughput under different number of hovering points.}
\label{fig_3040K}
\end{figure}

{\color{black}To verify the valid performance of our proposed algorithm, we have changed the number of hovering points in the experiment\footnote{{\color{black}We have also changed the distribution of hovering points. The newly added experimental results can be found in Appendix A.}}. We plot the average throughput value of DQN, QL, MCTS, respectively when $K=20, 30, 40$. As shown in Fig.~\ref{fig_3040K}, firstly, MCTS still exhibits the best performance among all the algorithms when the number of hovering points changes. Secondly, the average throughput grows as the number of hovering points $K$ increases because the action set of the UAV is expanded, and the UAV has more options of actions in the training process. Thus, the performance of all the algorithms improves when the number of hovering points increases.}


\section{Conclusions}
\label{sec5}
In this paper, we investigated a UAV-aided wireless system where the UAV provides task offloading services for mobile users. To maximize the average throughput, we have proposed an MCTS-based path planning scheme with the energy consumption and user fairness constraints. Then we have improved our algorithm in terms of the redundant training times. Experimental results demonstrated that the proposed algorithm has a better performance than DQN and QL, and the timesaving-MCTS algorithm can save about 30\% of training time than the original one.

{\color{black}In addition, our algorithm can also be applied to multi-user and multi-agent systems. For multi-agent systems, multiple instances of search tree can be run independently and their statistics combined to yield the final result. We will consider it as one of our future work.}

\begin{appendices}
\section{The effect of the distribution of hovering points}

\begin{figure}
\centering
\subfigure[average throughput with original distribution]{
  \includegraphics[width=0.23\textwidth]{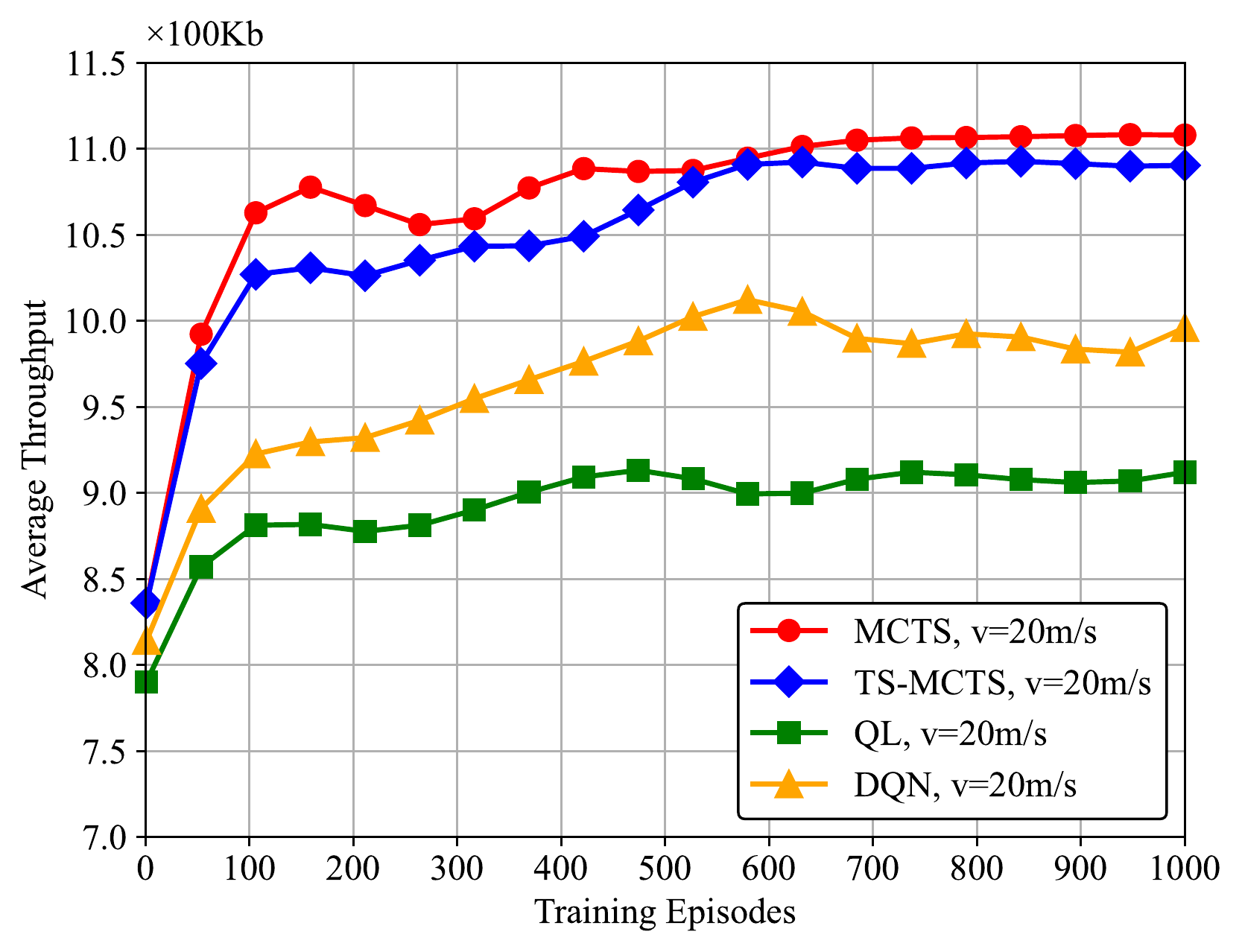}
}
\subfigure[average throughput with uniform distribution]{
  \includegraphics[width=0.23\textwidth]{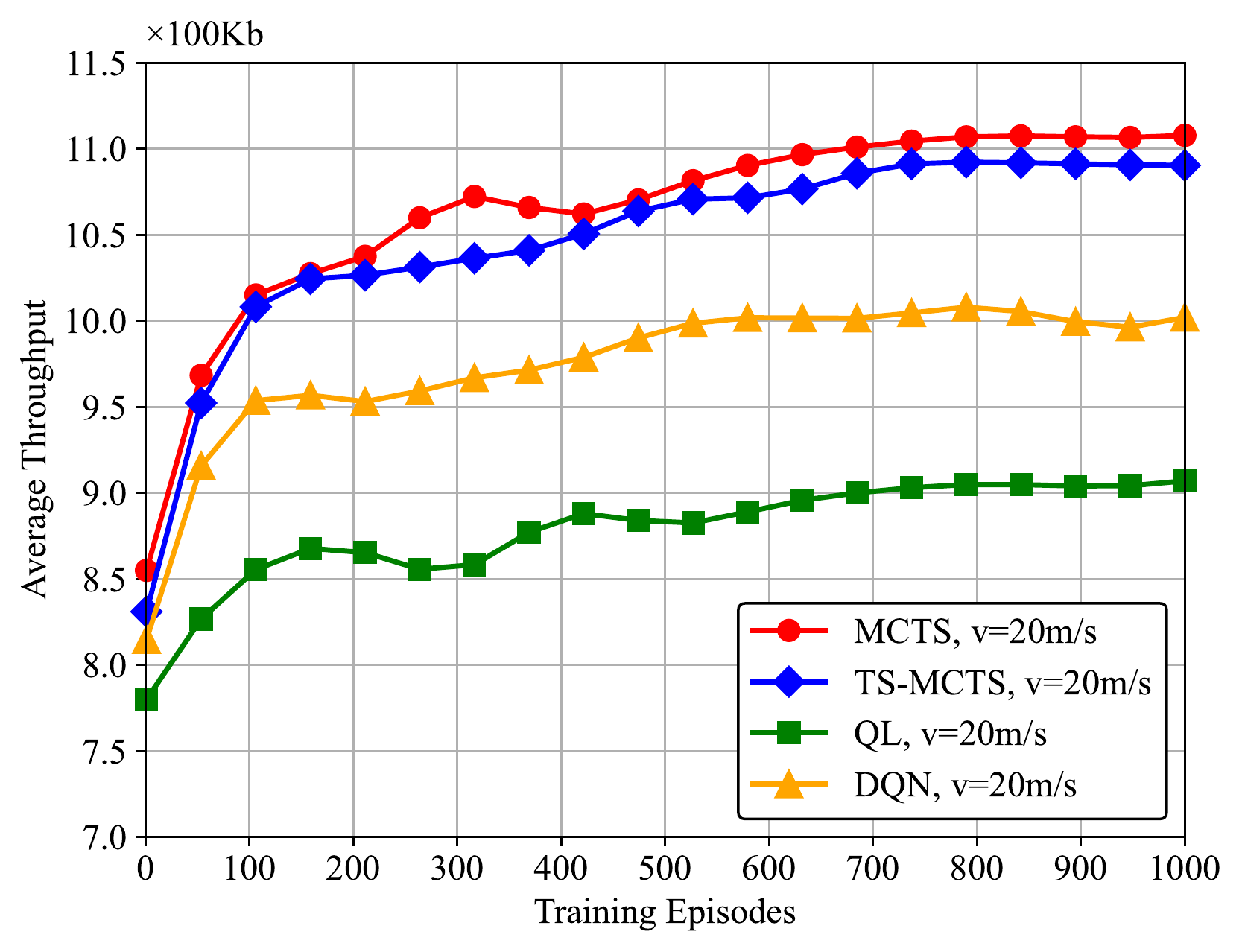}}
 \caption{The throughput performance with two different hovering points distributions.} \label{logr}
\label{uniform}
\end{figure}

To better show the effect of distribution of hovering points, we uniformly distribute 20 hovering points and compare the performance with the original one. The performances of these two distributions are illustrated in Fig.~\ref{uniform}.

As can be seen from these two figures, all the algorithms can achieve similar convergence values with two different hovering point distributions, which indicates that the random movement of users plays a more important role in achieved throughput. Due to the random movement of users, the specific distribution of hovering points has limited impacts on the achieved throughput. Besides, our proposed algorithm can achieve the largest average throughput under both the uniform distribution and the original distribution.

\section{A simple case of 3-D UAV trajectory}

We also consider a 3-D UAV trajectory. The hovering points are distributed on three planes with different heights, where $H_1$=25m, $H_2$=50m, $H_3$=75m, and there are 9 hovering points on each plane. Moreover, We have considered the users' fluctuant movement due to the varied topography. In this way, the users' positions are also set as 3-D coordinates. Following random way point model, the users can move both horizontally and vertically.

\begin{figure}
\centering
\includegraphics[height=5.5cm]{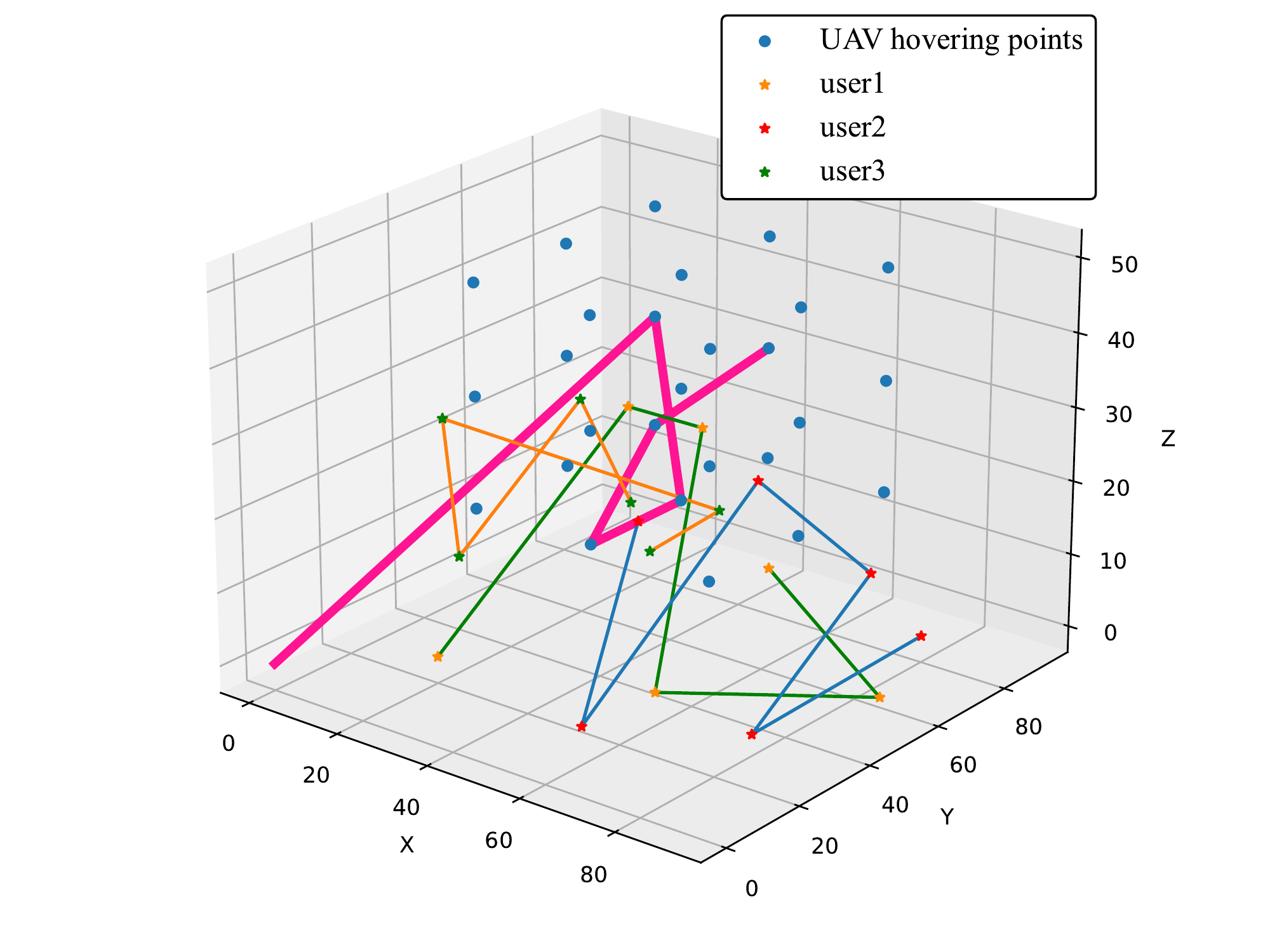}
\caption{The 3-D UAV trajectory.}
\label{3D_1}
\end{figure}

\begin{figure}
\centering
\includegraphics[height=4.3cm]{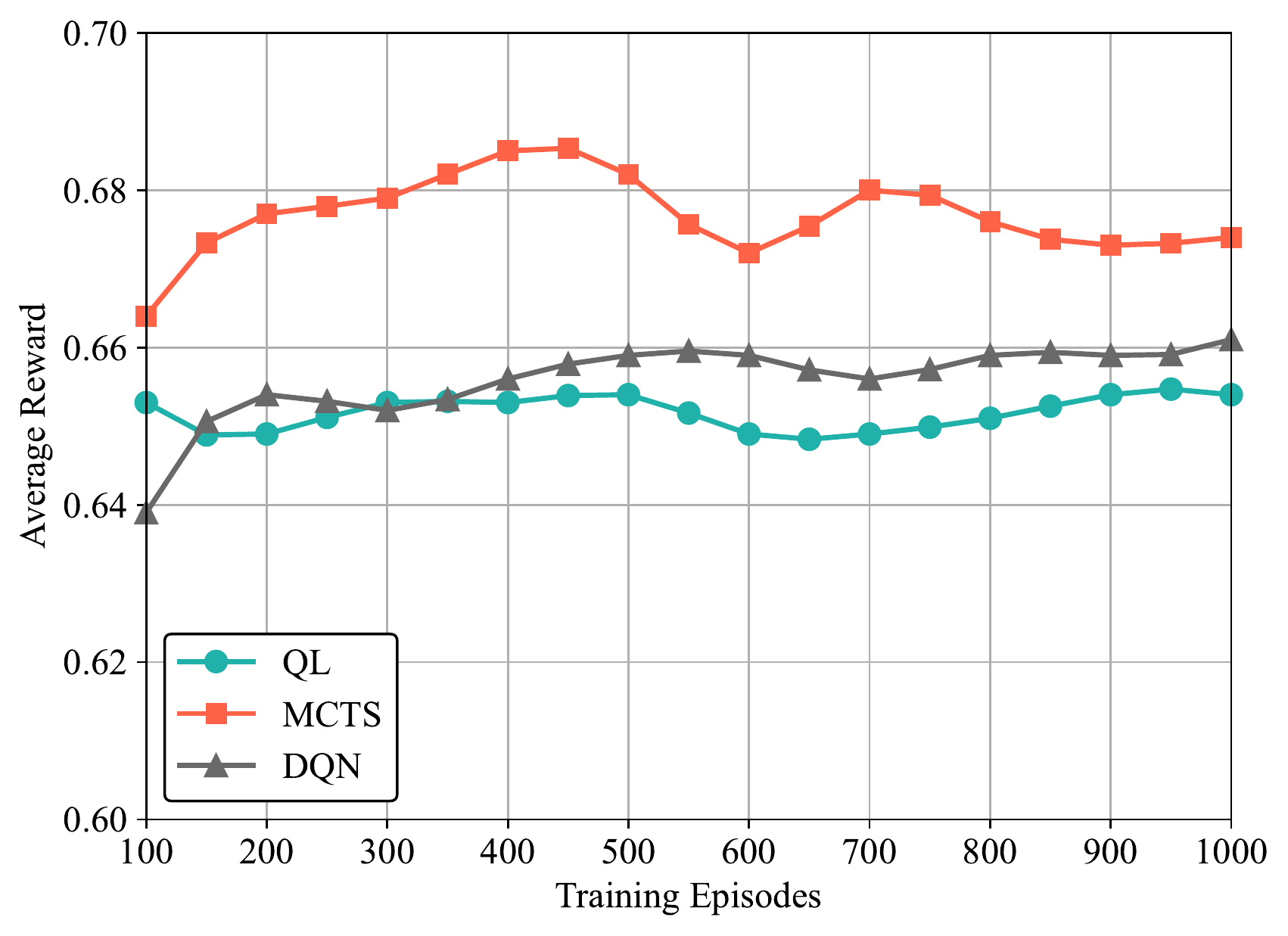}
\caption{The average reward value with 3-D UAV trajectory.}
\label{3D_2}
\end{figure}

Fig.~\ref{3D_1} illustrates the trajectory of the UAV and users with 3-D coordinates. In this figure, the pink line denotes the UAV trajectories, and the yellow line, blue line and green line denote the trajectories of three users, respectively. Fig.~\ref{3D_2} depicts the performance of average system reward with three different UAV path planning algorithms. As can be seen from this figure, MCTS-based path planning algorithm can obtain the largest average reward value among all the algorithms, which demonstrates the effectiveness of MCTS in 3-D trajectory optimization.

\end{appendices}

\bibliographystyle{IEEEtran}
\bibliography{1111}

\end{document}